\documentclass{elsart4}

\usepackage{amssymb}

\usepackage{graphics}
\usepackage{graphicx}
\usepackage{epsfig}
\usepackage{amssymb}
\journal{Physica B}

\begin{document}

\begin{frontmatter}


 \title{History dependent magnetoresistance in lightly doped La$_{2-x}$Sr$_{x}$CuO$_4$ thin films}
\author{Xiaoyan Shi, Dragana Popovi\'{c}\corauthref{cor1}}
\address{National High Magnetic Field Laboratory and Department of Physics, Florida State University, Tallahassee, Florida 32310, USA}
\author{C. Panagopoulos}
\address{Department of Physics, University of Crete and FORTH, GR-71003 Heraklion, Greece\\Division of Physics and Applied Physics, Nanyang Technological University, Singapore}
\author{G. Logvenov, A. T. Bollinger, I. Bo\v{z}ovi\'{c}}
\address{Brookhaven National Laboratory, Upton, New York 11973, USA}
 \corauth[cor1]{Corresponding author.  Tel: +1 850 644 3913; fax:+1 850 644 5038.
  {\it Email address}: ${\bf dragana@magnet.fsu.edu}$ (D. Popovi\'c).}





\begin{abstract}
The in-plane magnetoresistance (MR) in atomically smooth La$_{2-x}$Sr$_{x}$CuO$_4$ thin films grown by molecular-beam-epitaxy was measured in magnetic fields $B$ up to 9~T over a wide range of temperatures $T$.  The films, with $x=0.03$ and $x=0.05$, are insulating, and the positive MR emerges at $T<4$~K.  The positive MR exhibits glassy features, including history dependence and memory, for all orientations of $B$.  The results show that this behavior, which reflects the onset of glassiness in the dynamics of doped holes, is a robust feature of the insulating state.
\end{abstract}

\end{frontmatter}

\section{Introduction}
\label{}

Understanding the origin and the role of various nanoscale inhomogeneities observed in underdoped cuprates is one of the major open issues in the study of high-temperature superconductivity (HTSC).  The nature of the ground state at low charge carrier concentrations is of particular interest, because it is from this state that HTSC emerges with doping.  While the long-range antiferromagnetic (AF) order of the parent compound is already destroyed at very low level of hole doping (\textit{e.g.} $x\approx 0.02$ in La$_{2-x}$Sr$_{x}$CuO$_4$), short range AF correlations of the Cu spins persist \cite{Kastner98}.  In La$_{2-x}$Sr$_{x}$CuO$_4$ (LSCO), each of these AF domains has a weak ferromagnetic (FM) moment associated with it and oriented along the $c$ axis, \textit{i.e.} perpendicular to CuO$_2$ ($ab$) planes.  The direction of the FM moment is uniquely linked to the phase of the AF order \cite{thio88,thio90,lavrovSS01}.  At a relatively low temperature $T_{SG}$ ($T_{SG}$ -- spin glass transition temperature), the moments in different AF domains undergo cooperative freezing \cite{Cho92}, so that the ground state of Cu spins is the so-called cluster spin glass (SG).  The SG phase in LSCO emerges with the first added holes and extends all the way to slightly overdoped $x\simeq 0.19$ \cite{Niedermayer98,Christos-SG}.  On the other hand, several experiments suggest that, in lightly doped (nonsuperconducting) LSCO, charge is clustered in antiphase boundaries \cite{lavrovSS01,Matsuda1,Matsuda2,Wakimoto00} that separate the hole-poor AF domains in CuO$_2$ planes \cite{Julien99,Singer02NQR,Dumm03EM,Ando02Ranisotropy,Ando03MR}.  The nature of the charge ground state, however, has only recently attracted more attention.

In particular, in LSCO with $x=0.03$, which does not superconduct at any $T$, resistance noise spectroscopy \cite{Ivana-PRL,Ivana-PRB11} shows that, deep within the SG phase ($T\ll T_{SG}$), doped holes form a collective, glassy state of charge domains or clusters located in CuO$_2$ planes.  The results strongly suggest that glassy freezing of charges occurs as $T\rightarrow 0$.  These conclusions are supported by impedance spectroscopy \cite{Jelbert08}.  In the same $T$ range, both out-of-plane and in-plane magnetoresistance (MR) exhibit the emergence, at low fields, of a strong, positive component for all orientations of the magnetic field $B$ \cite{Ivana-PRL,Ivana-PRB10}.  The positive MR (pMR) grows rapidly with decreasing $T$ and thus dominates the MR in the entire experimental $B$ range at the lowest $T$.  At higher $T$ and $B$, on the other hand, the MR is negative.  The mechanism of the negative MR at high $T$ is attributed \cite{kotov07} to the reorientation of the weak FM moments.  Most strikingly, unlike the negative MR, the pMR reveals clear signatures of glassiness, such as hysteresis and memory \cite{Ivana-PRL,Ivana-PRB10,Ivana-Stripes}.  Similar behavior was observed in La$_2$Cu$_{1-x}$Li$_x$O$_4$ (Li-LCO) with $x=0.03$, where long-range AF order is still present \cite{Ivana-PRB10}.  This material, however, remains insulating for all $x$ \cite{kastner88}, and dielectric response provides evidence for slow \cite{Jelbert08,park05} and glassy \cite{park05} charge dynamics in Li-LCO at low $x$.  Detailed studies of the hysteretic and memory effects in the pMR of single crystals of both La$_{1.97}$Sr$_{0.03}$CuO$_4$ \cite{Ivana-PRL,Ivana-PRB11,Ivana-PRB10,Ivana-Stripes} and La$_{2}$Cu$_{0.97}$Li$_{0.03}$O$_{4}$ \cite{Ivana-PRB10,Ivana-LiLCOPRB} have provided strong evidence that such history dependent behavior reflects primarily the dynamics of doped holes.

It is clearly of great interest to investigate the evolution of this glassy charge state with doping.  For that purpose, LSCO films grown by molecular beam epitaxy (MBE) are particularly suitable because, in addition to their uniform thickness and precise crystal orientation, the doping can be controlled continuously and with high accuracy.  Obviously, it is necessary to establish first whether history dependent positive MR observed in single crystals is also present in MBE-grown films, and thus independent of the growth conditions.  Here we present a study of the low-$T$ magnetotransport in MBE-grown LSCO films with $x=0.03$ and $x=0.05$, which demonstrates the presence of the pMR and the associated glassy effects.

\section{Experiment}
\label{}

The LSCO films were grown by atomic-layer-by-layer molecular beam epitaxy (ALL-MBE) \cite{MBE} on LaSrAlO$_4$ substrates with the $c$ axis perpendicular to the surface. The films were deposited at $T\approx 680~^{\circ}$C under $3\times 10^{-6}$~Torr ozone partial pressure. The growth was monitored in real-time by reflection high energy electron diffraction (RHEED) which showed that the films were atomically smooth and without any secondary-phase precipitates. The films are 75 unit cells (about 100~nm) thick; the measured $c = 1.312$~nm. Finally, 160~nm of gold was evaporated \textit{in situ} on top of the films for contacts.  The films were patterned using UV photolithography and ion milling to fabricate Hall bar patterns with the length $L = 2$~mm and the width $W = 0.3$~mm. The distance between the voltage contacts is 1.01~mm, and their width is 50~$\mu$m.  In order to remove any excess oxygen, the films were subsequently annealed in high vacuum ($4-5\times 10^{-5}$~mBar) for over an hour at $200-250~^{\circ}$C.

The in-plane sample resistance $R$ was measured with a standard four-probe ac method ($\sim11$~Hz) in the Ohmic regime, at $T$ down to 0.3~K realized in a $^3$He cryostat.  In the MR measurements, the current $I\perp B$.

\section{Results and Discussion}
\label{}
\begin{figure}
\begin{center}
\includegraphics[width=6cm]{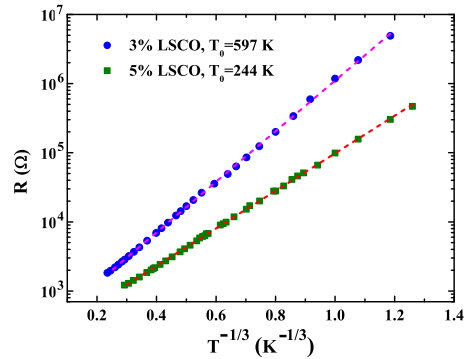}
\end{center}
\caption{The temperature dependence of the zero-field cooled in-plane resistance $R$: below 70~K for 3\% LSCO ($\bullet$), and below 40~K for 5\% LSCO ({\tiny $\blacksquare$}) samples.  Dashed lines are linear fits.}
\label{Fig.1}
\end{figure}

Both samples exhibit insulating behavior, with a two-dimensional (2D) form of the variable-range hopping (VRH) $R=R_{0}\exp(T_{0}/T)^{1/3}$ obeyed up to about 70~K in La$_{1.97}$Sr$_{0.03}$CuO$_4$ (3\% LSCO) and 40~K in La$_{1.95}$Sr$_{0.05}$CuO$_4$ (5\% LSCO), as shown in Fig.~\ref{Fig.1}.
The localization length $\xi$ obtained from the VRH fits~\cite{2DOS} is $\sim 40$~\AA\, in 3\% ($T_0=597$~K), and $\sim 60$~\AA\, in 5\% ($T_0=244$~K) LSCO.  Thus the doped holes are here localized more strongly than in a
high-quality LSCO single crystal \cite{sasagawa98} with a nominal doping of 3\%, where $\xi\sim 90$~\AA\, and 2D VRH is obeyed up to 30~K \cite{Ivana-PRB11,Ivana-PRB10,Ivana-Stripes}.  In both cases, the range of the data is sufficiently wide to allow a reliable determination of the VRH exponent and without the need to invoke a $T$ dependence of the prefactor $R_0$.
The exponent $1/3$, characteristic of 2D hopping, does not depend on doping, in contrast to early results on ceramic LSCO samples \cite{LSCO-hopping}, but in analogy with more recent studies of polycrystalline LSCO \cite{Hucker-hopping}.  Since the VRH conduction in both LSCO films and single crystals is dominated by 2D physics, at least in the experimental $T$ range (see also Ref.~\cite{kotov07}), the thickness of the films is not expected to affect the dimensionality of the transport properties.

\begin{figure}
\begin{center}
\includegraphics[width=6cm]{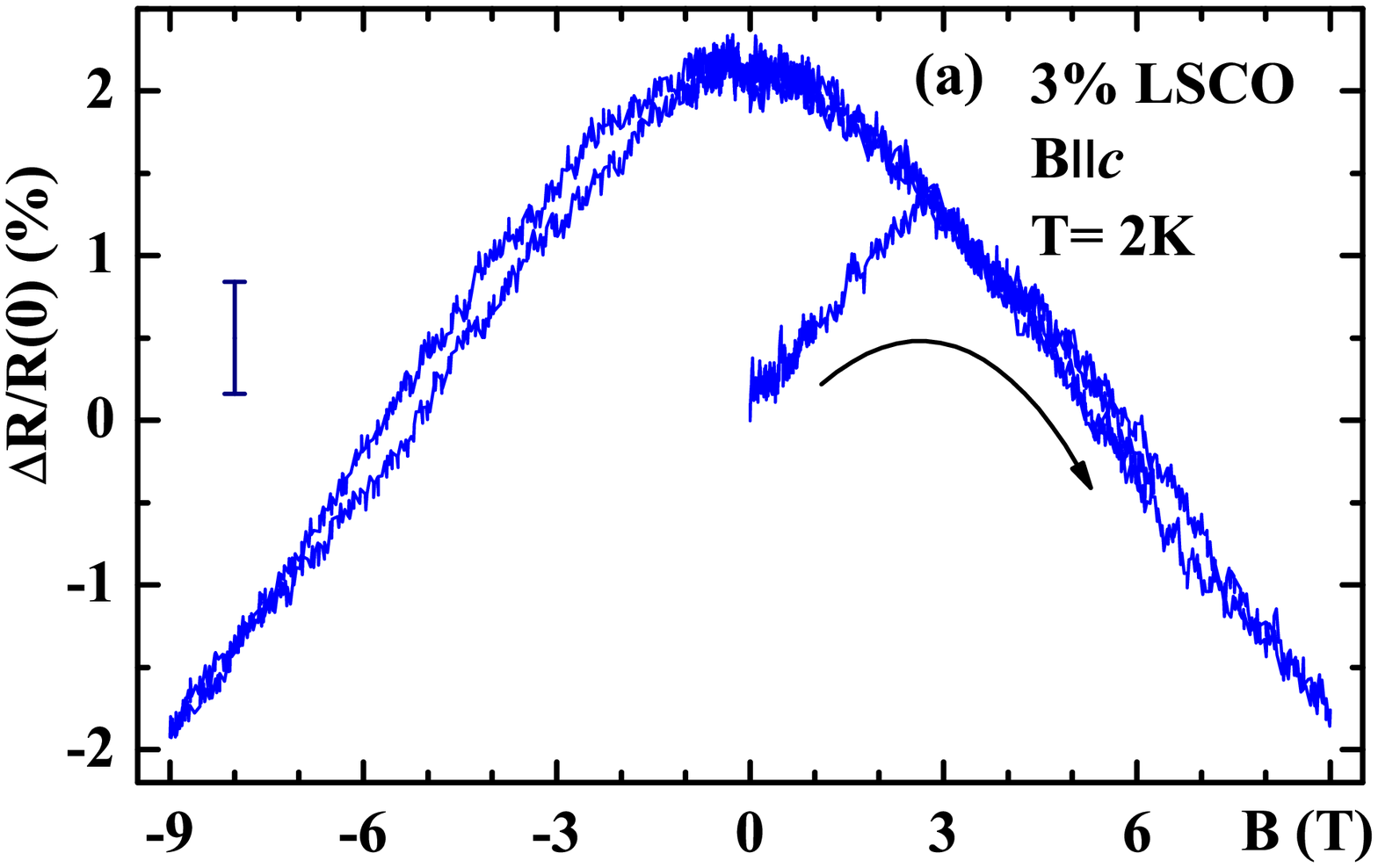}
\includegraphics[width=6cm]{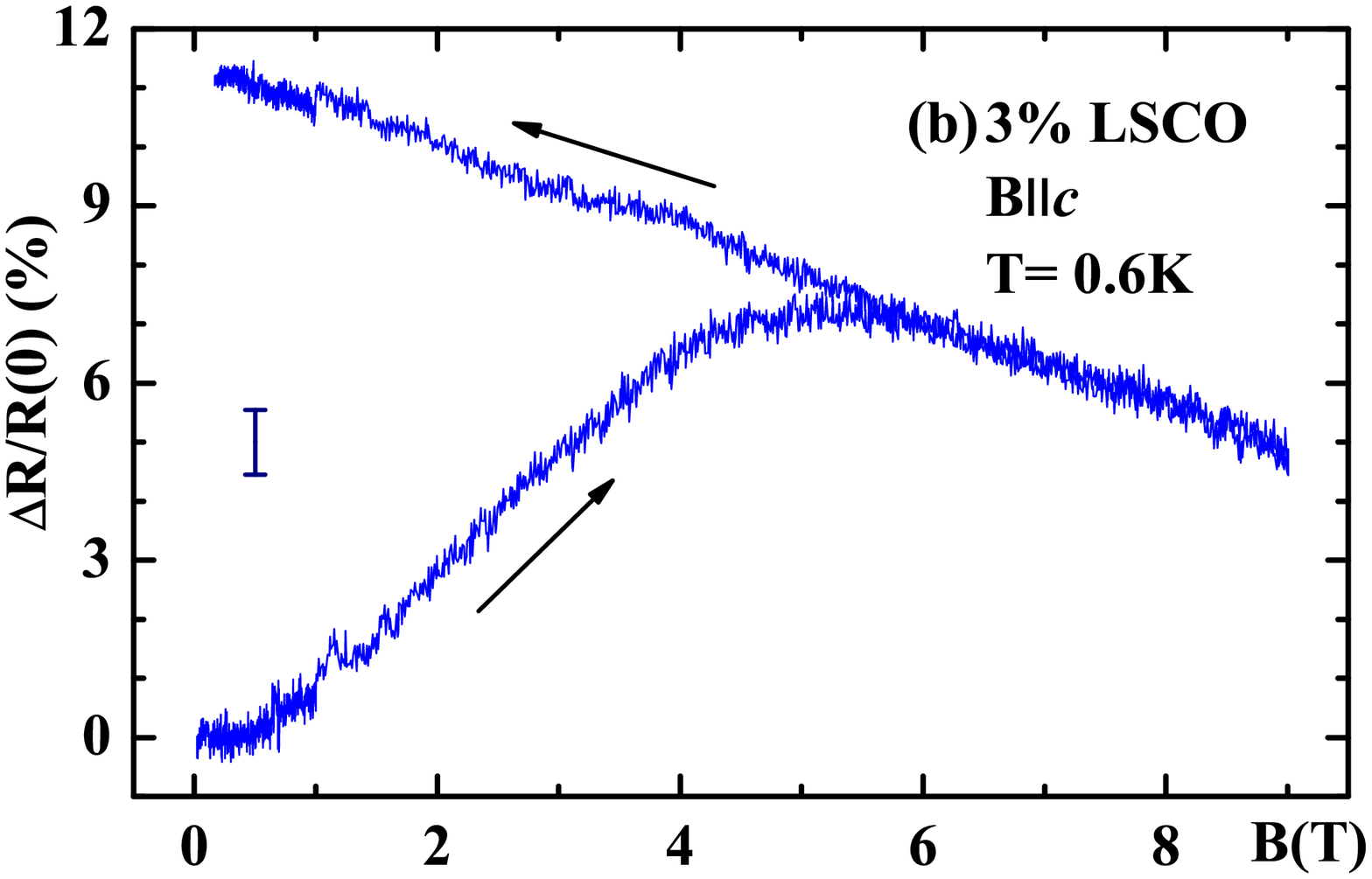}
\end{center}
\caption{3\% LSCO.  Hysteretic behavior of the positive component of the in-plane MR for $B\parallel c$ at (a) $T=2$~K, and (b) $T=0.6$~K.  The error bars correspond to the maximum change in the MR due to $T$ fluctuations (6~mK in (a), and 2~mK in (b)).  (a) The arrow denotes the direction of the initial sweep from 0 to 9~T.  This was followed by sweeps from 9~T to $-9$~T, then to 6~T, and then back to 0.  The first sweep is clearly different from subsequent ones, which are the same within the error.  (b) The arrows denote the direction of $B$ sweeps: from 0 to 9~T, then back to 0.  In both (a) and (b), the sweep rates were low enough (0.005~T/min for $B<1$~T, 0.02~T/min for $B>1$~T) to avoid the sample heating.}
\label{Fig.2}
\end{figure}

Similar to the behavior in $x=0.03$ La$_{2-x}$Sr$_{x}$CuO$_4$ and La$_2$Cu$_{1-x}$Li$_x$O$_4$ single crystals~\cite{Ivana-PRL,Ivana-PRB10}, at low enough $T$, the MR of both 3\% and 5\% LSCO films exhibits the emergence of the positive component at low magnetic fields $B$ for all field orientations.  Here the pMR appears for $T<3-4$~K, as illustrated in Fig.~\ref{Fig.2}(a) for 3\% LSCO with $B$ applied parallel to the $c$ axis.
As $T$ is reduced, the pMR increases in magnitude and dominates the MR over an increasingly large range of $B$ [Fig.~\ref{Fig.2}(b)].  History dependent behavior is present only in the $B$ region where the pMR was initially observed (Fig.~\ref{Fig.2}) after zero-field cooling.  If the applied $B$ is sufficiently large to lead to the negative MR, the MR will remain negative upon subsequent field sweeps (Fig.~\ref{Fig.2}).  In such a case, only the curve obtained in the first sweep will be different from the others, the latter being symmetric around $B=0$ [Fig.~\ref{Fig.2}(a)].  This result is also similar to the findings in heavily underdoped YBa$_2$Cu$_3$O$_{6+x}$ \cite{Ando-YBCO} at low $T$, where they were attributed to the freezing of the AF domains and their charged boundaries.

The important point in Fig.~\ref{Fig.2} is that $R$ does not return to its initial value after removing the magnetic field: the system acquires a memory.  In order to erase the memory in 3\% and 5\% LSCO films discussed here, it was necessary to warm them up to $T>10$~K.  This is somewhat different from La$_{1.97}$Sr$_{0.03}$CuO$_4$ and La$_2$Cu$_{0.97}$Li$_{0.03}$O$_4$ single crystals, where the memory could be erased by warming them up to only a $T$ where pMR vanishes ($\sim 1$~K in LSCO, $\sim 4$~K in Li-LCO, with $\ll T_{SG}\approx 7-8$~K in both materials) \cite{Ivana-PRB10,Ivana-Stripes,Ivana-PRL}.  The memory effects that are described in more detail below were obtained after thermal cycling first to $T>10$~K and then cooling to the measurement $T$.

\begin{figure}[h]
\begin{center}
\includegraphics[width=6cm]{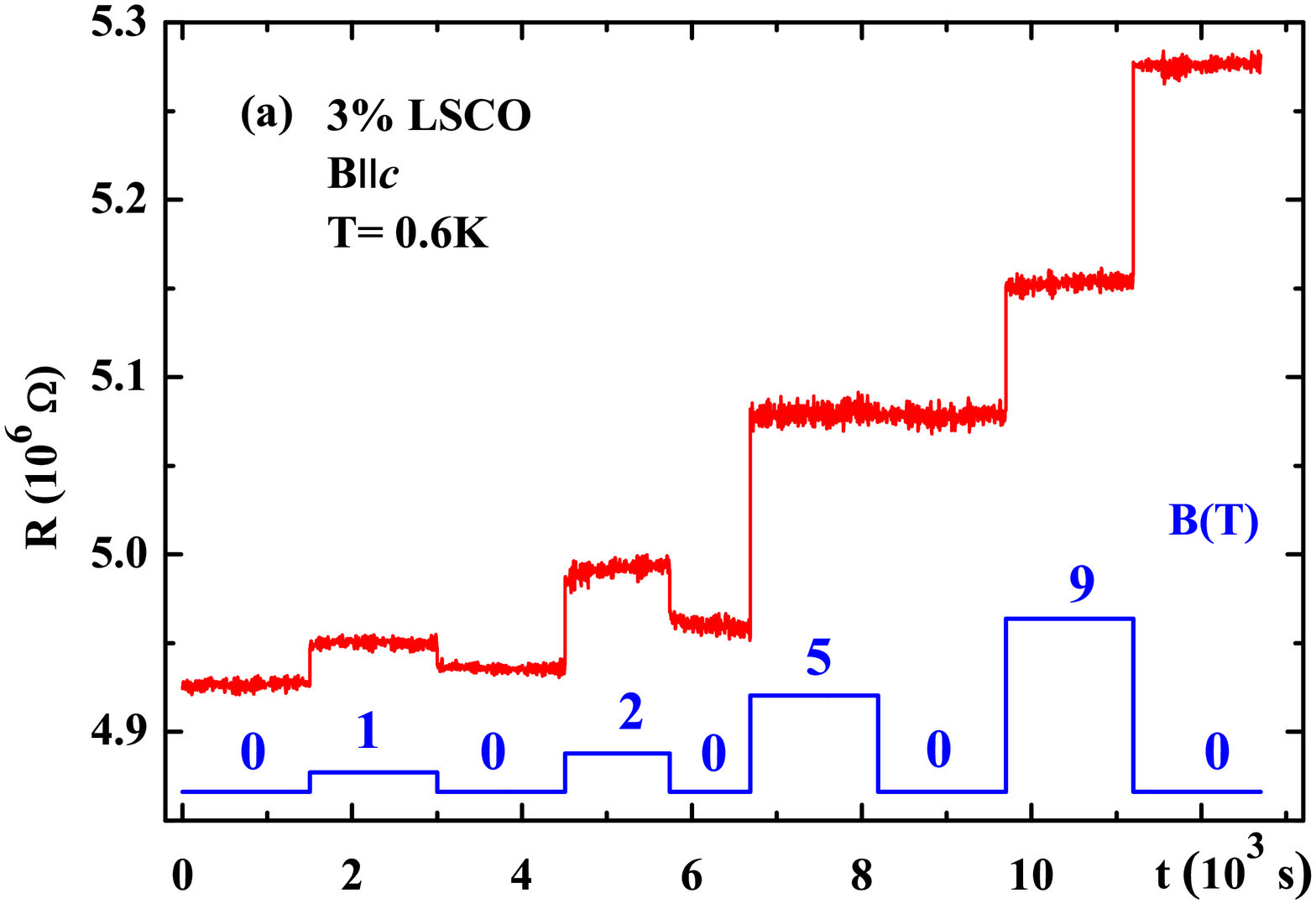}
\includegraphics[width=6cm]{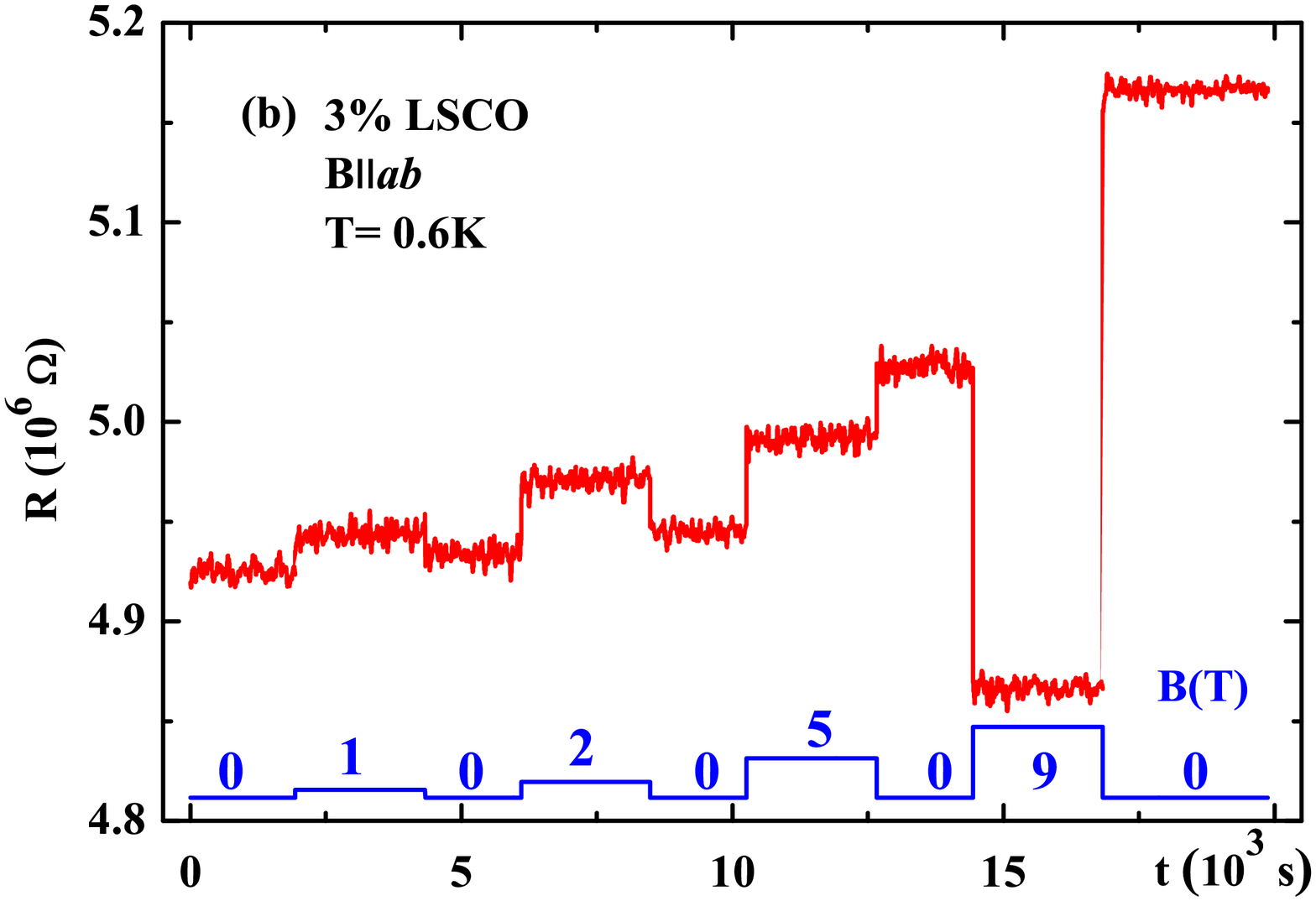}
\end{center}
\caption{3\% LSCO.  The in-plane resistance $R$ as a function of time upon the subsequent application and removal of $B=1, 2, 5, 9$~T at $T=0.6$~K for (a) $B\parallel c$, and (b) $B\parallel ab$.  The bottom traces in (a) and (b) show the experimental protocol.}
\label{Fig.3}
\end{figure}

\begin{figure}
\begin{center}
\includegraphics[width=6cm]{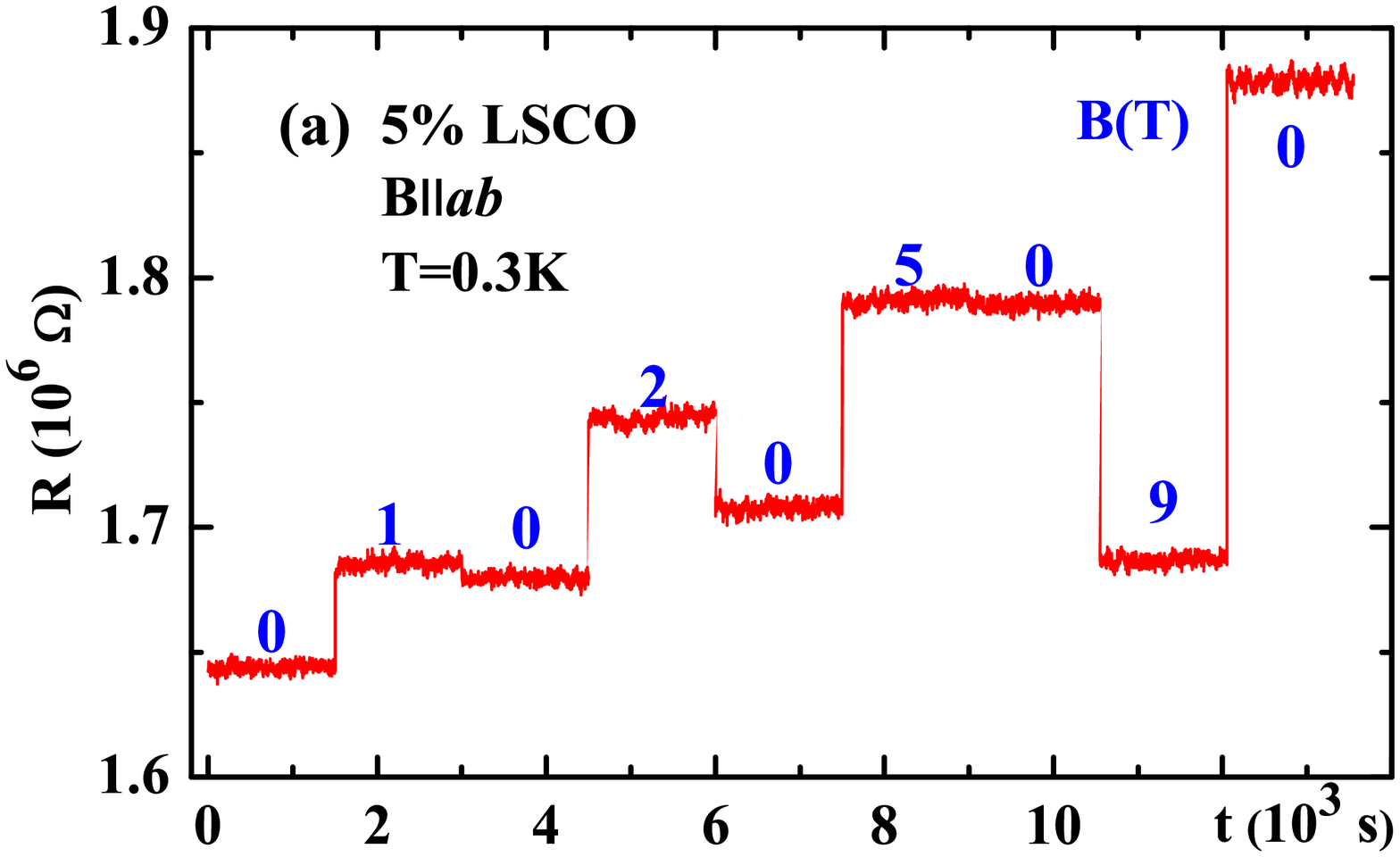}
\includegraphics[width=6cm]{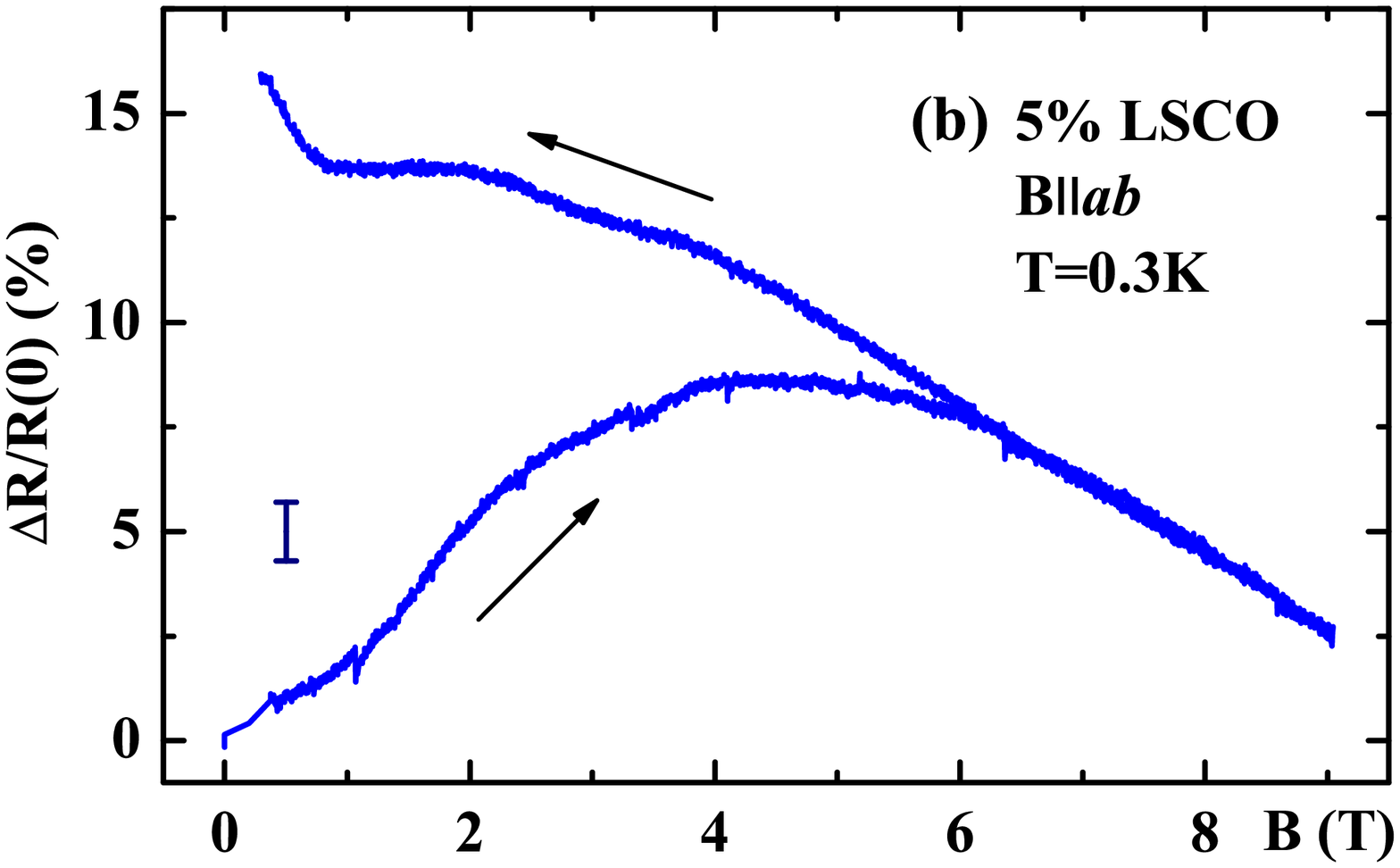}
\end{center}
\caption{5\% LSCO.  (a) The in-plane resistance $R$ as a function of time upon the subsequent application and removal of $B=1, 2, 5, 9$~T, as shown, at $T=0.3$~K for $B\parallel ab$.  (b) Hysteretic MR for the same $T$ and $B$ orientation.  The arrows denote the direction of $B$ sweeps.  The error bar corresponds to the maximum change in the MR due to temperature fluctuations of 1~mK.  The sweep rate was 0.01~T/min.}
\label{Fig.4}
\end{figure}

For example, Fig.~\ref{Fig.3}(a) shows the response of $R$ in 3\% LSCO to the subsequent application and removal of different $B$ values under the same conditions, $B\parallel c$ and $T=0.6$~K, as those in Fig.~\ref{Fig.2}(b).  The values of the zero-field resistance $R(B=0)$ clearly depend on
the magnetic history, in a manner consistent with the behavior of the MR [Fig.~\ref{Fig.2}(b)].  Qualitatively the same memory effects are observed also with the field oriented parallel to CuO$_2$ planes [Fig.~\ref{Fig.3}(b)].  Likewise, the resistance of the 5\% LSCO sample depends on its magnetic history, as illustrated in Fig.~\ref{Fig.4} for $B\parallel ab$ and $T=0.3$~K.

The hysteretic, low-$T$ positive MR described here is thus similar to that observed in insulating LSCO and Li-LCO single crystals and, therefore, independent of the sample growth conditions.  In all cases, history dependent effects emerge for both $B\parallel c$ and $B\parallel ab$.  Moreover, in the measurements with $B\parallel ab$, the current $I\perp B$ in the case of MBE-grown LSCO films, whereas $I\parallel B$ in the studies of single crystals.  Hence, there is no significant dependence on the $B$ orientation.  In single crystals, the pMR emerges at $T<T_{SG}$, and there is no hysteresis observed in the magnetization in the regime of the hysteretic pMR \cite{Ivana-LiLCOPRB}.  So far, it has not been possible to determine $T_{SG}$ in LSCO films independently, since it is notoriously difficult to measure the magnetization or magnetic susceptibility of thin films in general.  While the value of $T_{SG}$ in LSCO films remains an open question, it would be interesting to perform also other types of experiments, such as dielectric susceptibility measurements, on these films to gain further insight into the intriguing nature of the charge response in lightly doped cuprates.

\section{Summary}
\label{}

Low-$T$ studies of the in-plane magnetotransport in the insulating MBE-grown films of La$_{2-x}$Sr$_{x}$CuO$_4$ with $x=0.03$ and $x=0.05$ have revealed the emergence of the positive magnetoresistance and the associated history dependent effects.  This study confirms that this behavior, which reflects the glassy dynamics of doped holes, is a robust feature of the insulating state.  The next challenge is to track the fate of this state at higher doping, as the transition to a superconducting state is approached.\\

\noindent {\bf Acknowledgments}\\

This work was supported by NSF DMR-0905843, NHMFL via NSF
DMR-0654118, US DOE project MA-509-MACA, EURYI,
MEXT-CT-2006-039047, and the National Research Foundation, Singapore.




\begin{thebibliography}{00}




\bibitem{Kastner98} M. A. Kastner \textit{et al.},
Rev. Mod. Phys. \textbf{70}, 897 (1998).

\bibitem{thio88} T. Thio \textit{et al.},
Phys. Rev. B \textbf{38}, 905 (1988).

\bibitem{thio90} T. Thio \textit{et al.},
Phys. Rev. B \textbf{41}, 231 (1990).

\bibitem{lavrovSS01} A. N. Lavrov \textit{et al.},
Phys. Rev. Lett. \textbf{87}, 017007 (2001).

\bibitem{Cho92} J. H. Cho \textit{et al.},
Phys. Rev. B \textbf{46}, 3179 (1992).

\bibitem{Niedermayer98} C. Niedermayer \textit{et al.},
Phys. Rev. Lett. \textbf{80}, 3843 (1998).

\bibitem{Christos-SG} C. Panagopoulos \textit{et al.},
Phys. Rev. B \textbf{66}, 064501 (2002).

\bibitem{Matsuda1} M. Matsuda \textit{et al.},
Phys. Rev. B \textbf{62}, 9148 (2000).

\bibitem{Matsuda2} M. Matsuda \textit{et al.},
Phys. Rev. B \textbf{65}, 134515 (2002).

\bibitem{Wakimoto00} S. Wakimoto \textit{et al.},
Phys. Rev. B \textbf{62}, 3547 (2000).

\bibitem{Julien99} M.-H. Julien \textit{et al.},
Phys. Rev. Lett. \textbf{83}. 604 (1999).

\bibitem{Singer02NQR} P.~M. Singer \textit{et al.},
Phys. Rev. Lett. \textbf{88}, 047602 (2002).

\bibitem{Dumm03EM} M. Dumm \textit{et al.},
Phys. Rev. Lett. \textbf{91}, 077004 (2003).

\bibitem{Ando02Ranisotropy} Y. Ando \textit{et al.},
Phys. Rev. Lett. \textbf{88}, 137005 (2002).

\bibitem{Ando03MR} Y. Ando \textit{et al.},
Phys. Rev. Lett. \textbf{90}, 247003 (2003).

\bibitem{Ivana-PRL} I. Rai\v{c}evi\'{c} \textit{et al.},
Phys. Rev. Lett. \textbf{101}, 177004 (2008).

\bibitem{Ivana-PRB11} I. Rai\v{c}evi\'{c} \textit{et al.},
Phys. Rev. B \textbf{83}, 195133 (2011).

\bibitem{Jelbert08} G. R. Jelbert \textit{et al.},
Phys. Rev. B \textbf{78}, 132513 (2008).

\bibitem{Ivana-PRB10} I. Rai\v{c}evi\'{c} \textit{et al.},
Phys. Rev. B \textbf{81}, 235104 (2010).

\bibitem{kotov07} V. N. Kotov \textit{et al.},
Phys. Rev. B \textbf{76}, 224512 (2007).

\bibitem{Ivana-Stripes} I. Rai\v{c}evi\'{c} \textit{et al.},
submitted to J. Supercond. Nov. Magn. (2011).

\bibitem{kastner88} M. A. Kastner \textit{et al.},
Phys. Rev. B \textbf{37}, 111 (1988).

\bibitem{park05} T. Park \textit{et al.},
Phys. Rev. Lett. \textbf{94}, 017002 (2005).

\bibitem{Ivana-LiLCOPRB} I. Rai\v{c}evi\'{c}, 
D. Popovi\'c, C. Panagopoulos, L.~Benfatto, M. B. Silva Neto, E. S. Choi, T. Sasagawa
(unpublished).

\bibitem{MBE} I. Bozovic,
IEEE Trans. Appl. Supercon. \textbf{11}, 2686 (2001).

\bibitem{2DOS} For the 2D density of states,
we used $N_{2}\approx (0.7-4.5)\times 10^{15}$(cm$^2$\, eV)$^{-1}$ ($N=(2-13)$/(eV\, cell) \cite{DOS}).

\bibitem{DOS} R. L. Greene \textit{et al.},
Solid State Commun. \textbf{63}, 379 (1987).

\bibitem{sasagawa98} T. Sasagawa \textit{et al.},
Phys. Rev. Lett. \textbf{80}, 4297 (1998).

\bibitem{LSCO-hopping} B. Ellman \textit{et al.},
Phys. Rev. B \textbf{39}, 9012 (1989).

\bibitem{Hucker-hopping} M. H\"{u}cker \textit{et al.},
Phys. Rev. B. \textbf{59}, R725 (1999).

\bibitem{kotov07} V. N. Kotov \textit{et al.},
Phys. Rev. B \textbf{76}, 224512 (2007).

\bibitem{Ando-YBCO} Y. Ando \textit{et al.},
Phys. Rev. Lett. \textbf{83}, 2813 (1999).
%
\end{thebibliography}
\end{document}